\documentclass[12pt]{article}
\pdfoutput=1
\usepackage[square,comma,numbers,sort&compress]{natbib}
\usepackage{graphicx,epstopdf,amssymb,amsfonts,amsmath,amsthm,array,
mathrsfs,amscd}
\usepackage[vcentermath]{youngtab}
\usepackage{float}
\newcommand{\pbs}[1]{\let\temp=\\#1\let\\=\temp}
\numberwithin{equation}{section}
\def\be{\begin{equation}}\def\ee{\end{equation}}
\def\cvp{\raise 2pt\hbox{,}} 

 \def\tr{\mathop{\rm tr}\nolimits}

\def\re{\mathop{\rm Re}\nolimits}  

 \def\d{{\text d}} 
 \def\uN{\text{U}(N)}

\def\La{\Lambda}
\def\Tc{T_{\text c}}\def\Mc{M_{\text c}}

\def\npb#1#2#3{{\it Nucl.\ Phys.\ }{\bf B #1} (#2) #3}

\def\jhep#1#2#3{{\it J. High Energy Phys.\ }{\bf #1} (#2) #3}
\def\prd#1#2#3{{\it Phys.\ Rev.\ }{\bf D #1} (#2) #3}

\def\cmp#1#2#3{{\it Comm.\ Math.\ Phys.\ }{\bf #1} (#2) #3}

\def\imath#1#2#3{{\it Invent math }{\bf #1} (#2) #3}

\def\ahep#1#2#3{{\it Adv.\ High Energy Phys.\ }{\bf #1} (#2) #3}

\begin{document}
%
%
{\pagestyle{empty}
\begin{flushright} CERN-TH-2016-021 \end{flushright}
\parskip 0in
\

\vfill
\begin{center}
%
%

{\LARGE Black Hole Horizons}

\bigskip

{\LARGE and Bose-Einstein Condensation}

\vspace{0.4in}

Frank F{\scshape errari}
\\

\medskip

{
\it Service de Physique Th\'eorique et Math\'ematique\\
Universit\'e libre de Bruxelles (ULB) and International Solvay Institutes\\
Campus de la Plaine, CP 231, B-1050 Bruxelles, Belgique

\smallskip

\it Theoretical Physics Department\\
CERN, CH-1211 Gen\`eve, Suisse}

%

\smallskip
{\tt frank.ferrari@cern.ch}
\end{center}
\vfill\noindent

Consider a particle sitting at a fixed position outside of a stable black hole. If the system is heated up, the black hole horizon grows and there should exist a critical temperature above which the particle enters the black hole interior. We solve a simple model describing exactly this situation: a large $N$ matrix quantum mechanics modeling a fixed D-particle in a black hole background. We show that indeed a striking phenomenon occurs: above some critical temperature, there is a non-perturbative Bose-Einstein condensation of massless strings. The transition, even though precisely defined by the presence of the condensate, cannot be sharply detected by measurements made in a finite amount of time. The order parameter is fundamentally non-local in time and corresponds to infinite-time correlations.

\vfill

\medskip
%
\begin{flushleft}
\today
\end{flushleft}
%
\newpage\pagestyle{plain}
\baselineskip 16pt
\setcounter{footnote}{0}

}


\newpage
\section{\label{IntroSec} Introduction and set-up}

In classical General Relativity, the Principle of Equivalence implies that absolutely nothing special happens when an infalling local observer crosses a black hole horizon, at least when the black hole is much larger than the Planck length. The quantum version of this scenario remains, however, poorly understood and is at the basis of puzzling paradoxes (see e.g.\ the recent review \cite{HarlowRev} and references therein). A basic difficulty is that the known formulations of the quantum mechanics of black holes are based on holography, a framework in which the time associated with the quantum mechanical unitary evolution matches the asymptotic time of the black hole geometry. The infinite redshift factor at the horizon then prevents in principle the detection of the crossing of the horizon, at least in finite time. Other aspects of the physics associated with black hole horizons are much better understood, like the presence of a continuous spectrum and the relation between the quasi-normal behaviour and thermalization (see e.g.\ \cite{MaldaBH,FL} and the review \cite{HubenyRev}).

The aim of the present paper is to propose and study a simple framework in which the crossing of a black hole horizon by a localized probe can be precisely discussed. We consider a stable black hole in the presence of a probe particle at a given position in space. A string-inspired set-up for this situation consists of a large number $N$ of background D-branes describing the black hole and a probe D-brane playing the role of the particle. A simple Hamiltonian modeling such a brane system was considered by Iizuka and Polchinski in \cite{IP1,IP2},
\be\label{HIP} H =\tr\Bigl(\frac{1}{2}\Pi^{2} + \frac{1}{2}m X^{2}\Bigr) + M a^{\dagger}_{i}a^{i} + \frac{\mu}{N}a^{\dagger}_{i}a^{\dagger}_{j}a^{i}a^{j}+ m\sqrt{\frac{\La}{N}}\, a^{\dagger}_{i}X^{i}_{\ j}a^{j}\, .\ee
The $N\times N$ matrix operators $X$ and $\Pi$ satisfy the reality conditions $\smash{(X^{i}_{\ j})^{\dagger}}=\smash{X^{j}_{\ i}}$, $\smash{(\Pi^{i}_{\ j})^{\dagger}=\Pi^{j}_{\ i}}$ and the canonical commutation relations $\smash{[X^{i}_{\ j},\Pi^{k}_{\ l}] = i\delta^{i}_{l}\delta^{k}_{j}}$. They model the black hole degrees of freedom. The creation operators $a^{\dagger}$ satisfy $[a^{i},a^{\dagger}_{j}]=\delta^{i}_{j}$. They create open strings joining the probe D-particle and the black hole branes. The mass $M$ of these open strings is proportional to the distance between the black hole branes and the probe particle and thus parameterizes the position of the particle in space. The cubic coupling $a^{\dagger}Xa$ is the simplest non-trivial coupling between the black hole and the open strings one can imagine. The quartic coupling $a^{\dagger}a^{\dagger}aa$ ensures the stability, i.e.\ the existence of the partition function, of the model.\footnote{A high temperature analysis straightforwardly shows that the model is well-defined for all values of $\mu>\La/2$.}

Iizuka and Polchinski studied the model in the limit $M\rightarrow\infty$ and found the expected quasi-normal behaviour \cite{IP1,IP2}. Their results strongly support the idea that a simple Hamiltonian like \eqref{HIP} is able to capture important features of black hole physics, in spite of the fact that more realistic D-brane probe Hamiltonians include more interaction terms (see e.g.\ \cite{ferbranes} for explicit constructions). Similarly, we believe that the qualitative phenomena found in the present paper using \eqref{HIP} will persist, at least in a wide class of more complicated examples. For example, one may consider an arbitrary interacting potential $V(X)$ for the matrix $X$ instead of the simple harmonic potential, or include more general interaction terms like $a^{\dagger}X^{2}a$ in the Hamiltonian, without changing the basic features of our discussion below in any crucial way. 

The limit $M\rightarrow\infty$ studied in \cite{IP1,IP2} corresponds to a probe sitting infinitely far away from the black hole. The open strings are then infinitely long and massive and thus cannot be thermally excited. This simplifies the analysis considerably. However, for the purpose of studying the physics associated with the crossing of the horizon, it is clearly essential to put the probe at a finite distance from the black hole and thus to keep $M$ finite. This explains why the physics we describe below is not found in \cite{IP1,IP2}. 


The black hole interpretation of \eqref{HIP} suggests a surprising behaviour of the model as a function of temperature. If we gradually increase $T$, the entropy, or equivalently the area of the horizon, will grow.\footnote{We are dealing with a stable black hole and thus $\partial S/\partial T>0$.} It is then clear that there should exist a critical temperature $\Tc$ at which the black hole captures the probe particle. The high temperature regime $T>\Tc$ should then correspond to the particle probing the black hole interior! This heuristic picture suggests that something rather drastic has to happen \emph{above} some critical temperature for models of the form \eqref{HIP}. One could expect a sharp transition at $T=\Tc$, associated with the particle crossing the horizon.

This is rather paradoxical. Non-trivial phases are usually found by going to low temperature, not to high temperature. Moreover, the infinite redshift at the horizon suggests that massless degrees of freedom play an important role in understanding the transition from $T<\Tc$ to $T>\Tc$. This seems to contradict the standard lore in statistical physics which indicates that low energy excitations become irrelevant at high temperature. It hints at a sort of UV/IR relation for models of the form \eqref{HIP}, with important high temperature (UV) properties governed by massless (IR) modes. Even more puzzling, the infinite redshift seems to imply that the transition, whatever sharp, cannot be detected in finite time!

Our main result is to show that, indeed, the model \eqref{HIP} displays a new kind of large $N$ phase transition having all the expected properties explained above. The transition corresponds to a non-perturbative\footnote{The transition is not seen to any finite order of perturbation theory.} Bose-Einstein condensation of massless strings \emph{above} $\Tc$.\footnote{This is an example of a so-called \emph{inversed} phase transition, occurring in the unexpected temperature regime. For some values of the parameter, the transition can also be \emph{reentrant,} see below.} Even though the transition is sharply defined by the presence of the condensate, it is of infinite order and even analytic, in the sense that all finite time physical observables are analytic at $\Tc$. The order parameter is fundamentally non-local in time and corresponds to the infinite time limit of a two-point function. The limit is zero in the normal phase for which the particle is outside the black hole and non-zero in the condensed phase, after the crossing of the horizon. In the following, we explain how this interesting physics comes about in the simplest possible terms. A much more detailed presentation will be given in a companion paper \cite{BHBE2}.

To the best of our knowledge, the physics we find has not been discussed before, in spite of the fact that similar models have been studied in the literature, for example in the very interesting paper \cite{IKLL}. In particular, we find a sharp signal associated with the horizon, whereas previous works did not. The reason could be that previous calculations used rather uncontrolled approximations, which cannot capture the effect we describe.


%
\section{\label{TwoPtSec} Two-point function and spectral density}

The fundamental observable on which we focus is the real-time two-point function at finite temperature $T$,
\be\label{twopt1} C(t) = \frac{1}{N}\bigl\langle a^{i}(t)a^{\dagger}_{i}\bigr\rangle_{T} = \frac{1}{2\pi}\int_{-\infty}^{+\infty} \tilde C(\omega)e^{-i\omega t}\, \d\omega\ee
and its Euclidean counterpart
\be\label{twopt2} G(\tau) = \frac{1}{N}\bigl\langle a_{\text E}^{i}(\tau)a^{\dagger}_{i}\bigr\rangle_{T} =T\sum_{k\in\mathbb Z}G_{k} e^{-i\nu_{k}\tau}\, ,\ee
where $\nu_{k}=2\pi k T$ are the Matsubara frequencies and $a_{E}(\tau) = e^{\tau H}ae^{-\tau H}$ as usual. In \eqref{twopt2}, the first equality is valid for $0<\tau<1/T$ whereas the second equality extends the definition of the correlator for any value of Euclidean time. It is also very useful to introduce the \emph{spectral function}
\be\label{rhodef} \rho(\omega) = \frac{1}{2\pi}\bigl(1-e^{-\beta\omega}\bigr)\tilde C(\omega)\, ,\ee
in terms of which the normalized average number of open strings reads
\be\label{nform} n =\frac{1}{N}\bigl\langle a^{\dagger}_{i}a^{i}\bigr\rangle = \int_{-\infty}^{+\infty}\!\frac{\rho(\omega)}{e^{\beta\omega}-1}\,\d\omega \, .\ee
This relation, valid in the full interacting theory, formally coincides with the usual formula for the occupation number of an ideal Bose gas of spectral density $\rho$. Accordingly, the spectral function can be interpreted as giving the spectrum of open strings in our problem. In particular, if we turn off the couplings in \eqref{HIP}, $\rho(\omega)=\delta(\omega-M)$. Another useful object is the so-called resolvent,
\be\label{resdef} R(z) = \int_{-\infty}^{+\infty}\frac{\rho(\omega)}{z-\omega}\, \d\omega\, ,\ee
which is an analytic function on both the upper- and lower-half complex $z$-plane.

By using unitarity and the spectral decomposition of the two-point function, it is straightforward to prove the following important properties (see e.g.\ \cite{ARG} for a detailed account):

\noindent\ i) The spectral function $\rho$ is real, positive for $\omega\geq 0$, negative for $\omega\leq 0$ and normalized, $\int_{-\infty}^{+\infty}\!\rho(\omega)\,\d\omega = 1$.\footnote{The spectral density can have a negative tail for negative frequencies in the interacting theory at non-zero temperature because matrix elements of the form $\langle p| a^{\dagger}_{i}|q\rangle$ between energy eigenstates can be non-zero even if the associated energies satisfy $E_{p}<E_{q}$. This is explicitly seen in \cite{BHBE2}.}

\noindent\ ii) The Matsubara coefficients satisfy $G_{k}^{*}=G_{-k}$ and $\re G_{k}\geq 0$. In particular, $G_{0}\in\mathbb R^{+}$.

\noindent\ iii) The spectral function and the Matsubara coefficients \emph{at non-zero frequencies} can be obtained from the resolvent $R$,
\begin{align}\label{disceq} \rho(\omega) & = \frac{i}{2\pi}\lim_{\epsilon\rightarrow 0+}\bigl(R(\omega+i\epsilon)-R(\omega-i\epsilon)\bigr) + \beta n_{0}\omega\delta(\omega)\, ,\\
\label{GkReq} G_{k} &= -R(i\nu_{k}) + \beta n_{0}\delta_{k,0}\, .
\end{align}
The zero-frequency contributions \emph{cannot} be obtained from $R$ alone in general and are parameterized by a non-negative constant $n_{0}$. The constant $n_{0}$ is zero to all orders of perturbation theory, but will play a crucial role in our analysis.\footnote{If $Z=\tr e^{-\beta H}$ is the partition function and $\{|p\rangle\}$ denotes an orthonormal basis of eigenvectors of the Hamiltonian, $H|p\rangle = E_{p}|p\rangle$, then $n_{0}=\frac{1}{N Z}\sum_{p,q, E_{p}=E_{q}}e^{-\beta E_{p}}\sum_{i}|\langle p|a^{i}|q\rangle|^{2}$. From this expression, it is easy to prove that $n_{0}=0$ in perturbation theory. Moreover, $n_{0}$ cannot be obtained from $R$ because the spectral decomposition of $R$ only involves matrix elements $\langle p|a^{i}|q\rangle$ for $E_{p}\not = E_{q}$.}

\section{\label{SolSec} The solution of the model}

As explained in \cite{BHBE2}, the solution of the model in the large $N$ limit (thermodynamic potentials, real-time and Euclidean correlators) is fully encoded in the coefficients $G_{k}$ or, equivalently, in the function $\rho$. The $G_{k}$ themselves can be computed from an infinite hierarchy of equations which is derived in \cite{BHBE2}. It turns out that the solution of these equations greatly simplifies if
\be\label{approx} m\ll T\, .\ee
This limit is particularly interesting, because it captures the existence of the critical temperature $\Tc$ and the nature of the transition while simplifying the mathematics enormously. We thus focus on \eqref{approx} in the present letter, leaving the detailed discussion of the general case to \cite{BHBE2}.

The condition \eqref{approx} implies that the black hole is highly excited and can be treated classically. The finite temperature matrix quantum mechanics then reduces to a standard zero-dimensional matrix model which can be solved using well-known techniques \cite{MM}. The upshot is that the matrix $X$ can be replaced by a static density of eigenvalues, for example the Wigner semi-circle law in the case of the harmonic potential used in \eqref{HIP}.\footnote{Clearly, this discussion could be straightforwardly generalized to any potential of the form $\tr V(X)$.} To obtain the solution, one must also take care of the quartic, vector-like coupling in \eqref{HIP}. At large $N$, this can be easily done by introducing a Hubbard-Stratonovich auxiliary field $\phi$ as usual. 

Overall, the solution takes the following simple form. The resolvent \eqref{resdef} can be directly read off from Wigner semi-circle law,
\be\label{Rsol} R(z) =\frac{1}{2\La T}\Bigl[z-M_{*}-\sqrt{(z-M_{+})(z-M_{-})}\Bigr]\, ,\ee
where 
\be\label{Mpmdef} M_{\pm}=M_{*}\pm 2\sqrt{\La T}\ee
and $M_{*}$ is a ``renormalized mass'' fixed by the large $N$ saddle point equation associated with $\phi$,
\be\label{Msteq} M_{*} = M + 2\mu n\, ,\ee
$n$ being given by \eqref{nform}. 

The interpretation of this solution is quite simple. First, the effect of the vector-like quartic coupling in \eqref{HIP} is to renormalize the mass $M$  to $M_{*}$. If $M$ is large (particle far away from the black hole), very few strings will be excited, $n\ll 1$ and $M_{*}\simeq M$. When $M$ is decreased, the excitation number grows and $M_{*}>M$. Second, the open string spectrum is broadened around $\omega=M_{*}$ due to the thermal coupling to the black hole. The smooth part of the spectral function, given by the discontinuity of the resolvent across its branch cut as in \eqref{disceq}, is
\be\label{smoothrho} 
\rho_{\text{smooth}}(\omega) = \frac{1}{2\pi\La T}\sqrt{(M_{+}-\omega)(\omega-M_{-})} \quad \text{for}\quad M_{-}\leq\omega\leq M_{+}\, .\ee
The width $M_{+}-M_{-}=4\sqrt{\La T}$ of the broadening is of the order of the coupling energy between the open strings and the black hole. Let us note that if one waives the condition \eqref{approx}, the form of the spectral function $\rho$ is much more complicated, but the basic features are maintained \cite{BHBE2}: there is a smooth continuous spectrum, mainly centered on $M_{*}$, with a width growing with the temperature.

A last important comment on the solution is that the zero-frequency piece in the spectral function, or equivalently the Matsubara zero mode $G_{0}$, is not immediately fixed by the above considerations. Indeed, \emph{the term $\beta n_{0}\omega\delta(\omega)$ does not contribute to the resolvent and is thus not described by the Wigner's law.} We are going to discuss this term in details below.

\section{\label{BESec} The Bose-Einstein condensation}
%


A crucial consequence of the solution \eqref{smoothrho} is the existence of a lower bound on $M_{*}$, or equivalently on $M_{-}$,
\be\label{fundbound} M_{-}=M_{*}-2\sqrt{\La T}\geq 0\, .\ee
The problem with $M_{-}<0$ can be understood from the formula \eqref{nform}, which would predict $n=\infty$, a ``firewall.'' But this is excluded from unitarity, since it would correspond to a positive spectral density for negative frequencies. Equivalently, $R(0)$ and thus the Matsubara zero mode $G_{0}$, given by \eqref{GkReq}, would become complex, again violating unitarity. When the bound \eqref{fundbound} is saturated, massless strings appear in the spectrum, since $M_{-}$ is the lower end of the support of $\rho$. \emph{The obvious interpretation is thus that $M_{-}=0$ corresponds to the black hole horizon.} 

\begin{figure}
\centerline{\includegraphics[width=6in]{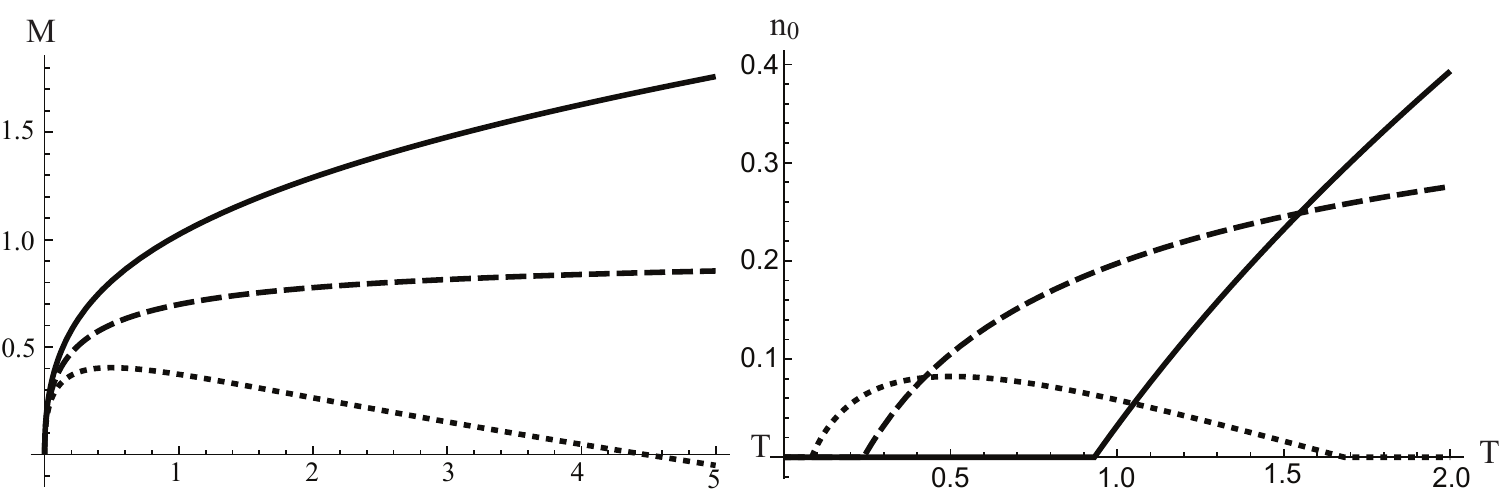}}
\caption{We set $\La=1$. Left inset: the critical line for $\mu=0.75$ (plain line), $\mu=1$ (dashed line) and $\mu=1.25$ (dotted line). Right inset: the Bose-Einstein condensate for the same values of $\mu$ and $M=1,0.5,0.3$ respectively.
\label{fig2}}
\end{figure}

To better understand the situation, it is useful to introduce the line of critical points $(\Tc,\Mc)$ defined by the conditions $M_{-}=0$, $n_{0}=0$. The explicit equation for the critical line can be found from \eqref{Msteq} and \eqref{nform} and reads
\be\label{TcMc} \Mc = 2\sqrt{\La\Tc}-\frac{16\mu}{\pi}\int_{0}^{1}\frac{\sqrt{x(1-x)}}{e^{4x\sqrt{\La/\Tc}}-1}\,\d x\, .\ee
The critical line is depicted on Fig.\ \ref{fig2}, left inset. One can distinguish three qualitatively different regimes, depending on the large $\Tc$ behaviour of $\Mc$, which is given by $\Mc = 2(\La-\mu)\sqrt{\Tc/\La} +\mu + O(1/\sqrt{\Tc})$.

Let us first discuss the case $\La/2<\mu<\La$. We fix $M>0$ and study the evolution of the spectral density as a function of $T$. At low temperature, one can reliably use perturbation theory, because the effective interaction energy $\sqrt{\La T}$ between the open strings and the black hole is very small. One then finds that $n_{0}=0$ and
\be\label{Mats1} G_{0} =-R(0)= \frac{1}{2\La T}\biggl[M_{-}+2\sqrt{\La T}-\sqrt{M_{-}\bigl(M_{-}+4\sqrt{\La T}\bigr)}\biggr]\, .\ee
When the temperature increases from zero, a straightforward analysis of the equation \eqref{Msteq} shows that $M_{-}$ decreases monotonically. Fig.\ \ref{fig2} shows that for some critical temperature $\Tc(M)$, we cross the critical line. At this point, $M_{-}=0$. This is the minimal possible value and around the critical point, $M_{-}\propto (T-\Tc)^{2}$. Plugging this into \eqref{Mats1}, we see that when $T>\Tc(M)$ the correct formula for the Matsubara zero mode as a function of $M_{-}$ is given by the analytic continuation,
\be\label{Mats2} G_{0} = \frac{1}{2\La T}\biggl[M_{-}+2\sqrt{\La T}+\sqrt{M_{-}\bigl(M_{-}+4\sqrt{\La T}\bigr)}\biggr]\quad\text{for}\ T>\Tc\, .\ee
It is important to stress that there is no such phenomenon of analytic continuation for the Matsubara coefficients $G_{k}=-R(i\nu_{k})$ at non-zero frequencies. Indeed, their branching points as a function of $M_{-}$ are situated at the complex values $M_{-}=i\nu_{k}$, which are never realized. Similarly, the expression \eqref{Rsol} of $R(z)$ as a function of $M_{-}$ is not changed when one goes through the transition for any non-zero complex $z$, and thus for any $z$ by analyticity in $z$.\footnote{This is also a consequence of the fact that $R$ is entirely fixed in terms of the $G_{k}$ for $k\geq 1$ by Carlson's theorem, see e.g.\ \cite{Cuniberti, ARG}.} This has a dramatic consequence: one no longer has $G_{0}=-R(0)$ and Eq.\ \eqref{GkReq} implies that 
\be\label{BEcondensate} n_{0} = \frac{1}{\La}\sqrt{M_{-}\bigl(M_{-}+4\sqrt{\La T}\bigr)}\not = 0 \quad\text{when}\ T>\Tc\, .\ee
Physically, this corresponds to a \emph{Bose-Einstein condensation of massless open strings.} In particular, for $T>\Tc$, 
\be\label{nformBE} n = n_{0}+ \int_{-\infty}^{+\infty}\!\frac{\rho_{\text{smooth}}(\omega)}{e^{\beta\omega}-1}\,\d\omega\ee
with a non-zero $n_{0}$. The condensate as a function of temperature is depicted on Fig.\ \ref{fig2}, right inset.\footnote{The transition from the normal phase to the condensed phase is somewhat reminiscent of the transition from the Coulomb branch to the Higgs branch on a D-brane moduli space, but this analogy must be taken very carefully. Coulomb and Higgs branches of the moduli space exist at finite $N$ and correspond to supersymmetric (and thus zero temperature) configurations. To the contrary, the phenomenon we describe depends \emph{crucially} on both the non-zero temperature and the large $N$ limit.}

The other cases $\mu=\La$ and $\mu>\La$ can be discussed in a similar way, according to the shape of the critical line depicted in Fig.\ \ref{fig2}. The important point is that the system is in the normal phase $n_{0}=0$ above the line and in the condensed phase $n_{0}\not = 0$ below the line. For example, for $\mu>\La$ and a small enough mass $M>0$, the system starts in the normal phase at low temperature, then goes to the condensed phase for some interval of temperature, and goes back to the normal phase at high temperature, a so-called \emph{reentrant} behaviour.

We have chosen to discuss the temperature dependence at fixed $M$, in order to emphasize the striking phenomenon of Bose-Einstein condensation \emph{above} some critical temperature. Of course, we can study what happens when $M$ varies at fixed $T$ as well. Physically, this corresponds to the probe particle moving in the fixed black hole background. From the shape of the critical line, we see that at large $M$ (particle far away from the black hole) we are always in the normal phase, as expected, whereas below some critical value of $M=\Mc(T)$, which is interpreted as corresponding to the location of the horizon, we are in the condensed phase. Moreover, one finds that for any value of $M<\Mc$, there is an associated value of $M>\Mc$ which yields the same $M_{-}$. This means that the continuous spectra of thermal strings precisely match for these two different values of $M$, as if the particle entering the black hole were actually bouncing off the horizon! The only difference between the two regions is the condensate which builds up more and more when $M$ is decreased.

A last fundamental point to mention is that all the Matsubara coefficients, including the zero mode, are \emph{analytic} (and thus infinitely differentiable) at the transition as a function of temperature. The same is true for all the thermodynamic potentials and finite-time correlation functions, which can be expressed in terms of the $G_{k}$ \cite{BHBE2}.\footnote{This is unlike the usual Bose-Einstein condensation found in weakly coupled Bose gas and is reminiscent of the Kosterlitz-Thouless phase transition.} This implies that, in spite of the sharp definition in terms of the Bose-Einstein condensate $n_{0}$, the transition cannot be detected in finite time! However, the condensate drastically affects the long-time behaviour of the correlation function \eqref{twopt1}, which goes to zero in the normal phase but goes to $n_{0}\not = 0$ in the condensed phase. The infinite-time correlations thus play the role of the order parameter.\footnote{Note that $\langle a^{i}\rangle=0$ for all $T$, including when a vanishingly small $\uN$-symmetry breaking term $\eta a^{i}+\text{H.c.}$ is introduced in the Hamiltonian. The expectation value $\langle a^{i}\rangle$ thus cannot play the role of an order parameter either.}

\section{\label{ConcSec} Conclusion}

By studying a simple quantum model for a D-particle probing a black hole background, we have discovered a new non-perturbative large $N$ phase transition which is naturally associated with the crossing of the black hole horizon. The basic physics is quite simple. When the temperature increases, the mass spectrum of open strings joining the particle and the black hole broaden and eventually massless strings can appear at a critical temperature $\Tc$.\footnote{Strictly speaking, this is always true only when $\La/2<\mu<\La$ in our model. See the discussion in the previous section for the general case.} In the dual geometric description, this corresponds to the growth of the horizon until it captures the particle. Above $\Tc$, the massless strings form a Bose-Einstein condensate. The transition is of infinite order (actually analytic) and cannot be detected in finite time. The infinite time limit of a two-point function plays the role of the order parameter. 

We believe that the phenomena just described are rather generic for Hamiltonians of the form \eqref{HIP}. It will be interesting and important to further explore their role in black hole physics and, more generally, to assess their possible relevance in other strongly coupled many-body quantum systems.

\subsection*{Acknowledgments}

I would like to thank Tatsuo Azeyanagi, Dongsu Bak, Adel Bilal, \'Edouard Br\'ezin, Michael Douglas, Paolo Gregori, Eliezer Rabinovici and Soo-Jong Rey for interesting discussions. I would also like to thank the Simons Center for Geometry and Physics in Stony Brook, USA (program ``Large $N$ Limit Problems in K\"ahler Geometry'') and the Asian Pacific Center for Theoretical Physics in Pohang, South Korea (workshop ``Liouville, Integrability and Branes (11)''), where the main results of the present paper were first presented.

This research in supported in part by the Belgian Fonds National de la Recherche Scientifique FNRS (convention IISN 4.4503.15, CDR grant J.0088.15 and MS grant) and the Advanced ARC project ``Holography, Gauge Theories and Quantum Gravity.''

\end{document}